\documentclass[12pt,preprint]{aastex}

\shorttitle{Evidence for Steady Heating}
\shortauthors{Warren, Winebarger, \& Brooks}

\begin{document}


\title{Evidence for Steady Heating: Observations of an Active Region Core with
  \textit{Hinode} and \textit{TRACE}}

\author{Harry P. Warren\altaffilmark{1}, Amy R. Winebarger\altaffilmark{2}, David
  H. Brooks\altaffilmark{1,3}}

\altaffiltext{1}{Space Science Division, Naval Research Laboratory, Washington, DC 20375}

\altaffiltext{2}{Department of Physics, Alabama A\&M, 4900 Meridian Street, Normal, AL
  35762}

\altaffiltext{3}{College of Science, George Mason University, 4400 University Drive,
  Fairfax, VA 22030}


\begin{abstract}
  The timescale for energy release is an important parameter for constraining the coronal
  heating mechanism. Observations of ``warm'' coronal loops ($\sim 1$\,MK) have indicated
  that the heating is impulsive and that coronal plasma is far from equilibrium.  In
  contrast, observations at higher temperatures ($\sim 3$\,MK) have generally been
  consistent with steady heating models.  Previous observations, however, have not been
  able to exclude the possibility that the high temperature loops are actually composed of
  many small scale threads that are in various stages of heating and cooling and only
  appear to be in equilibrium.  With new observations from the EUV Imaging Spectrometer
  (EIS) and X-ray Telescope (XRT) on \textit{Hinode} we have the ability to investigate
  the properties of high temperature coronal plasma in extraordinary detail. We examine
  the emission in the core of an active region and find three independent lines of
  evidence for steady heating. We find that the emission observed in XRT is generally
  steady for hours, with a fluctuation level of approximately 15\% in an individual
  pixel. Short-lived impulsive heating events are observed, but they appear to be
  unrelated to the steady emission that dominates the active region.  Furthermore, we find
  no evidence for warm emission that is spatially correlated with the hot emission, as
  would be expected if the high temperature loops are the result of impulsive heating.
  Finally, we also find that intensities in the ``moss,'' the footpoints of high
  temperature loops, are consistent with steady heating models provided that we account
  for the local expansion of the loop from the base of the transition region to the
  corona. In combination, these results provide strong evidence that the heating in the
  core of an active region is effectively steady, that is, the time between heating events
  is short relative to the relevant radiative and conductive cooling times.
\end{abstract}



\section{Introduction}

The coronal heating problem is one of the most fundamental open questions in solar
physics. The formation of the high temperature solar atmosphere is clearly related to the
magnetic fields generated in the solar interior, but how this magnetic energy is converted
to the thermal energy of the corona remains unknown. One important constraint on the
coronal heating mechanism is the time scale for energy release. If energy is deposited
into magnetic flux tubes on timescales that are very short compared to a characteristic
cooling time, then the corona will be filled with loops that are close to equilibrium and
appear to be steady. Active region observations have generally suggested that high
temperature loops are consistent with steady heating
\citep[e.g.,][]{porter1995,kano1995}. These studies have found that the observed evolution
of the emission is much slower than the radiative and conductive cooling timescales. There
has also been some success in modeling entire active regions with steady heating models
\citep{schrijver2004,warren2006b,winebarger2008,lundquist2008}. Finally,
\cite{antiochos2003} have argued that observations of the ``moss,'' the footpoints of high
temperature loops, are also consistent with steady heating. They found that the average
moss intensities are typically constant over many hours and loops cooling through 1\,MK
were not observed in the moss region they observed.

There is considerable evidence that coronal loops observed at lower temperatures ($\sim
1$\,MK) are evolving and not in equilibrium
\citep[e.g.,][]{aschwanden2001b,winebarger2003b,cirtain2007,urra2009}. These ``warm''
loops appear to have apex densities that are much higher than can be accounted for by
steady heating models \citep[e.g.,][]{winebarger2003,aschwanden2008b}. The properties of
the warm loops are more consistent with impulsive heating models
\citep[e.g.,][]{warren2002b,spadaro2003,warren2003}.

If it is true that hot loops are close to equilibrium while the warm loops are generally
heated impulsively then the coronal heating mechanism becomes even more difficult to
understand. The cooling time for short, hot loops is relatively rapid (typically a few
hundred seconds), so heating events on these loops would need to occur very
frequently. The cooling time for the warm loops is much longer (typically a few thousand
seconds), so impulsive heating events on these loops would be infrequent.

It is tempting to conjecture that the emission at high temperatures is also consistent
with impulsive heating models and that the apparent steadiness of this emission is the
result of the superposition of many evolving stands along the line of sight
\citep[e.g.][]{cargill1997,cargill2004,patsourakos2006}. Previous observations have not
been able to exclude this possibility. For example, it could be that the cooling loops are
not easy to detect in an active region core because they are faint relative to both the
bright moss and the extended corona in which the active region is embedded.

The EUV Imaging Spectrometer (EIS) and the X-ray Telescope (XRT) on the \textit{Hinode}
mission provide a new opportunity to observe active region emission in unprecedented
detail. EIS observes emission covering a very broad range of coronal temperatures: between
\ion{Fe}{7} and \ion{Fe}{24}, only 4 ionization stages of Fe are not present in the EIS
data (\ion{Fe}{18}--\textsc{xxi}). EIS also observes \ion{Ca}{14}, \textsc{xv},
\textsc{xvi}, and \textsc{xvii}, providing excellent coverage of the critical temperature
range around 3\,MK \citep{warren2008c}. XRT is a broadband imaging telescope that observes
high temperature plasma very efficiently and at high spatial resolution. The high cadence
XRT data complement EIS, which often observes at a much lower cadence, and allows us to
track the evolution of coronal plasma over a large field of view.

In this paper we use EIS and XRT observations to examine the properties of coronal plasma
in the core of an active region. The active region that we have selected (NOAA active
region 10960) is unusual in that most of the overlying warm loops are located to the north
and south of the region, providing a largely unobstructed view of the active region core
over a wide range of temperatures (see Figure~\ref{fig:context}). High cadence EIS
observations of this region, which was observed by \textit{Hinode} during the period 4--13
June 2007, have shown that the \ion{Fe}{12} 195.119\,\AA\ intensities, Doppler shifts, and
non-thermal widths in the moss are constant over long periods of time, suggesting steady
heating \citep{brooks2009b}. In this study we find three additional lines of evidence that
also indicate that the heating of high temperature loops is steady. We examine the
evolution of the emission in individual pixels in XRT and find that the vast majority of
the emission is constant to within approximately 15\% during this period. The inspection
of emission at different temperatures in the core of the active region shows that there is
no evidence for loops cooling from high temperatures. We find no relationship between the
warm emission (\ion{Fe}{10}--\ion{Fe}{14}) and the steady hot emission
(\ion{Ca}{14}--\ion{Ca}{17}). Finally, we find that we can bring all of the observed moss
intensities into agreement with steady heating models, if we allow for loop constriction
at the base of the loop. Previous modeling work had found significant discrepancies at the
lowest temperatures observed with EIS \citep{warren2008}.

None of these observations is conclusive; they do not necessarily exclude alternative
models that involve non-equilibrium processes. However, these observations do provide very
strong constraints on the mechanism responsible for producing the high temperature
emission observed in solar active regions.

\section{Observations}

The EIS instrument on \textit{Hinode} is a high spatial and spectral resolution imaging
spectrograph. EIS observes two wavelength ranges, 171--212\,\AA\ and 245--291\,\AA, with a
spectral resolution of about 22\,m\AA\ and a spatial resolution of about 1\arcsec\ per
pixel. There are 1\arcsec\ and 2\arcsec\ slits as well as 40\arcsec\ and 266\arcsec\ slots
available. The slit-slot mechanism and the CCD both image an area on the Sun 1024\arcsec\
in height, but a maximum of 512 pixels on the CCD can be read out at one time. Solar
images can be made using one of the slots or by stepping one of the slits over a region of
the Sun. Telemetry constraints generally limit the spatial and spectral coverage of an
observation. See \cite{culhane2007} and \cite{korendyke2006} for more details on the EIS
instrument.

\begin{figure}[t!]
\centerline{\includegraphics[clip,scale=1.0]{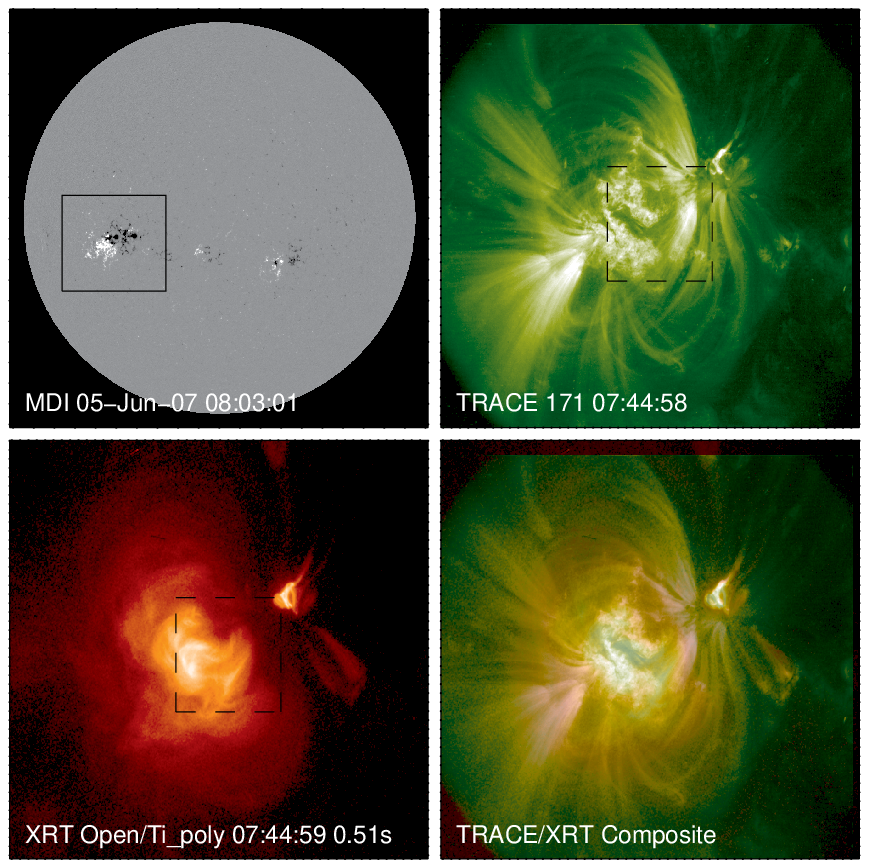}}
\caption{Context images for observations of NOAA active region 10960 taken on 5 June
  2007. Top left panel: An MDI/\textit{SoHO} magnetogram with the TRACE and XRT fields of
  view indicated. Top right and bottom left panels: TRACE 171\,\AA\ and XRT Ti-Poly images
  with the EIS field of view indicated. Bottom right panel: A composite TRACE and XRT
  image showing the difference in morphology between the hot and warm emission. Animations
  of the TRACE and XRT data are available with the electronic version of the manuscript
  (movie1\_trace.mpg, movie1\_trace\_moss.mpg and movie2\_xrt.mpg).}
\label{fig:context}
\end{figure}

For these observations the 1\arcsec\ slit was stepped over the central part of the active
region and 25\,s exposures were taken at each position. An area of
$128\arcsec\times128\arcsec$ was imaged in about 57 minutes (see
Figure~\ref{fig:context}). The observing sequence for this observation returned the
complete wavelength range from each detector so all of the lines observable with EIS are
potentially available in these data.

\begin{figure*}[t!]
\centerline{\includegraphics[clip,scale=1.0]{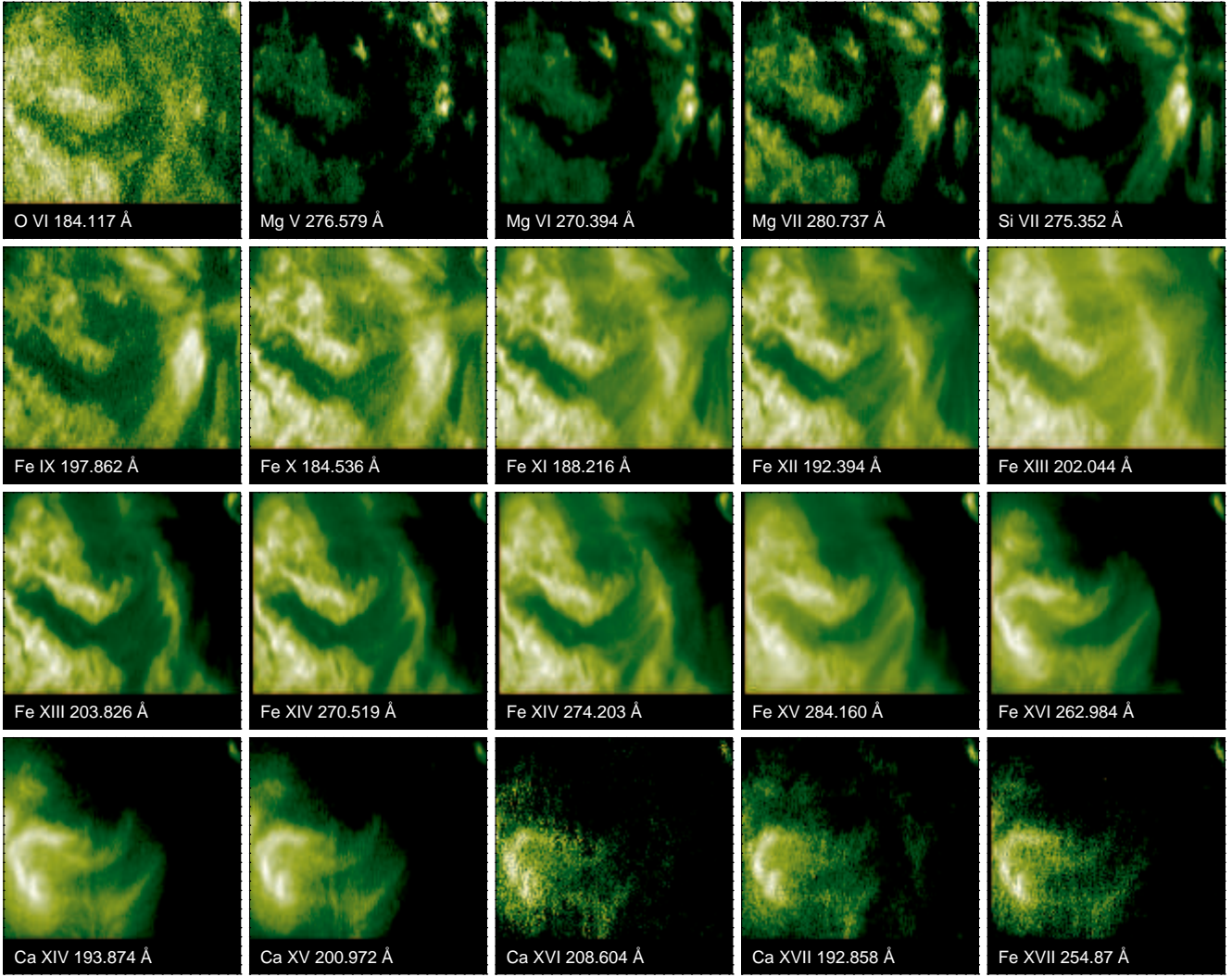}}
\caption{EIS rasters from the core of NOAA active region 10960 taken on 5 Jun 2007 between
  06:48:20 and 07:44:59 UT. With the exception of \ion{Ca}{17} 192.858\,\AA, which was
  deconvolved from a blend with other lines using a procedure described in the text, the
  intensities were derived from simple single or double Gaussian fits to the line
  profile.}
\label{fig:eis}
\end{figure*}

The raw data were processed using \verb+eis_prep+ to remove dark current, warm pixels, and
other instrumental effects using standard software. During the processing the observed
count rates are converted to physical units. Intensities from the processed data are
computed by fitting the observed line profiles with Gaussians. One line of particular
interest that requires special attention is \ion{Ca}{17} 192.858\,\AA, which is strongly
blended with \ion{Fe}{11} 192.813\,\AA\ and an \ion{O}{5} multiplet near 192.90\,\AA. An
algorithm for deconvolving this blend using \ion{Fe}{11} 188.216\,\AA\ and multi-component
fitting has been developed and is discussed in detail by \cite{ko2009}. We use this
approach here to determine the \ion{Ca}{17} intensities. The uncertainties for this line
are significantly greater than for the other lines that we consider. However, as
illustrated by the EIS rasters shown in Figure~\ref{fig:eis}, there is a strong similarity
between the deconvolved \ion{Ca}{17} 192.858\,\AA\ image and other rasters from ions
formed at similar temperatures, such as \ion{Ca}{15} 200.972\,\AA\ and \ion{Fe}{17}
254.87\,\AA.

The XRT on \textit{Hinode} is a high cadence, high spatial resolution (approximately
1\arcsec\ pixels) grazing incidence telescope that images the Sun in the soft X-ray and
extreme ultraviolet wavelength ranges. Temperature discrimination is achieved through the
use of focal plane filters. Because XRT can observe the Sun at short wavelengths, XRT
images can observe high temperature solar plasma very efficiently. The thinner XRT filters
allow longer wavelength EUV emission to be imaged and extend the XRT response to lower
temperatures. Further details on XRT are given in \cite{golub2007}.

The XRT data taken around the time of the EIS raster consisted of relatively high cadence
Open/Ti-Poly images with Open/Al-thick, Open/Be-thick, and G band images interleaved at a
cadence of about 600\,s. For this analysis we have processed all of the Open/Ti-Poly
images with \verb+xrt_prep+ to remove the dark current and to do exposure time
normalization. All of the images have been co-aligned to the initial frame using cross
correlation. To investigate the longer term evolution of this region we have considered
all of the XRT data within $\pm2$ hours of the start of the EIS raster. An example XRT
image is given in Figure~\ref{fig:context}. An animation of these data is available in the
electronic version of the manuscript.

In this analysis we also make reference to observations from \textit{TRACE}, the
\textit{Transition Region and Coronal Explorer} \citep{handy1999}. \textit{TRACE} is a
normal incidence, multi-layer telescope. There are 3 channels for imaging the corona:
\ion{Fe}{9}/\textsc{x} 171\,\AA, \ion{Fe}{12} 195\,\AA, and \ion{Fe}{15} 284\,\AA. There
is also a long wavelength channel for imaging the photosphere, chromosphere, and
transition region. All of the images are projected onto a single CCD and images in
different wavelengths must be taken sequentially. For these observations the data are
mostly from the 171\,\AA\ channel, with a few white light, 1600\,\AA, and 284\,\AA\ images
taken for context. The 171\,\AA\ data were processed with \verb+trace_prep+, despiked, and
co-aligned using cross-correlation. An example TRACE 171\,\AA\ image is given in
Figure~\ref{fig:context}. An animation of the data is available in the electronic version
of the manuscript.

\section{Analysis}

In this section we presented a detailed analysis of the EIS, XRT, and \textit{TRACE} data
available for this period. The EIS data represent a snapshot of the active region
accumulated over about an hour and yield little insight into the temporal evolution of the
emission. The XRT and TRACE observations, in contrast, have excellent temporal coverage
but somewhat limited diagnostic capabilities. In combination they allow us to establish a
detailed understanding of the plasma properties in the core of an active region and how
they evolve in time.

\subsection{Temporal Evolution}

\begin{figure*}[t!]
\centerline{\includegraphics[clip,scale=0.58]{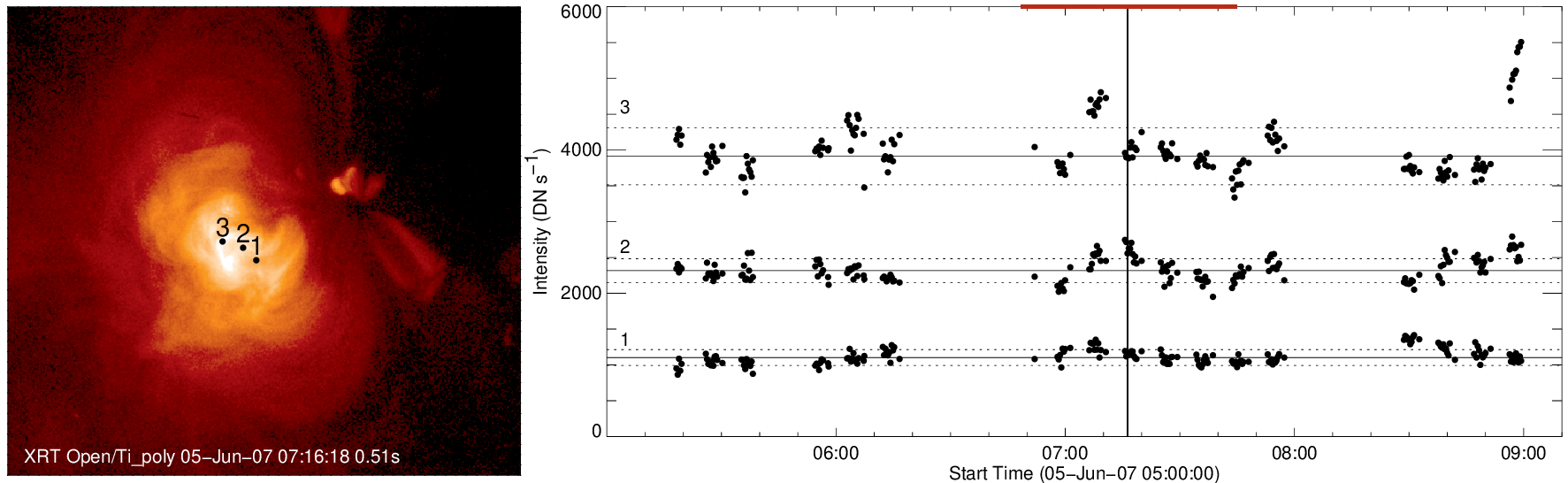}}
\centerline{\includegraphics[clip,scale=0.58]{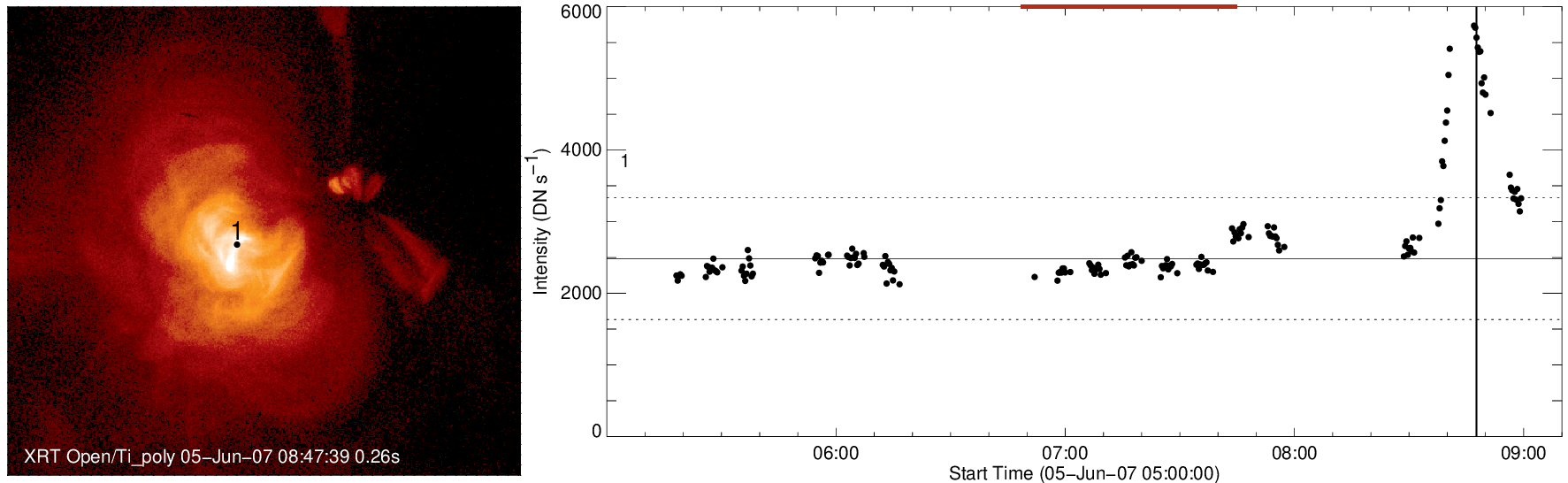}}
\centerline{\includegraphics[clip,scale=0.58]{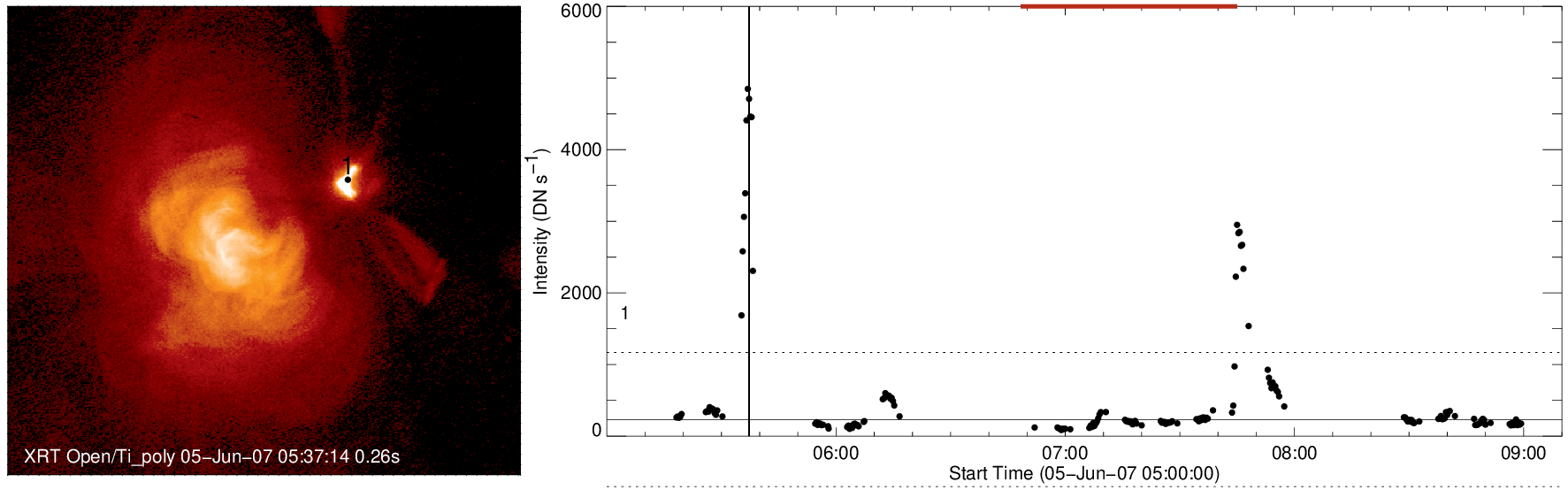}}
\caption{The evolution of the high temperature emission measured at various points in the
  active region core. The left panels show a single image from the XRT data set with the
  points indicated.  The right panels show the light curves for these points. For each
  point the median intensity ($\bar{I}$) is indicated with a solid line and
  $\bar{I}\pm\sigma_I$ is indicated by the dashed lines. The red bar at the top of panel
  highlights the time of the EIS raster. The vertical line indicates the time of the XRT
  image displayed on the left. The top panel illustrates relatively constant emission that
  dominates the core of the active region. The bottom panels illustrate the evolution of
  more variable emission.}
\label{fig:xrt}
\end{figure*}

As is suggested by the four-hour XRT movie associated with Figure~\ref{fig:context}, the
large scale pattern of the soft X-ray emission from this region is remarkably steady. To
illustrate this we have selected several points from the core of the active region and
plotted light curves for the emission in these individual pixels. The light curves, shown
in Figure~\ref{fig:xrt}, generally show fluctuation levels of about 10--15\% around the
median intensity for this time interval. In Figure~\ref{fig:xrt} we also show two points
that illustrate the evolution of transient loops. The lifetimes of these brightenings are
generally on the order of 1000\,s or less and show that the intensities in the core of the
active region are steady on timescales that are long compared to a cooling time.

\begin{figure}[h!]
\centerline{\includegraphics[clip,scale=0.95]{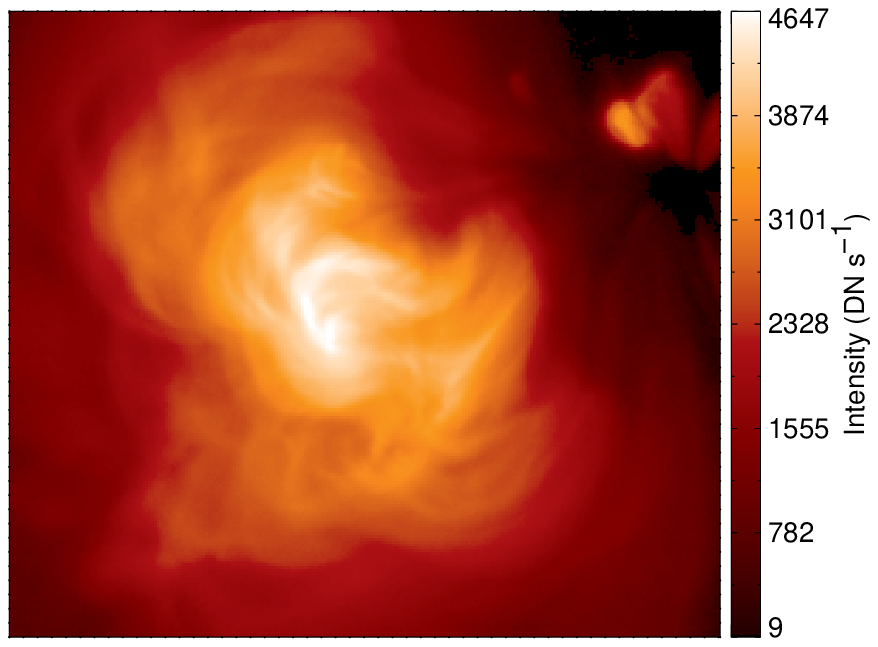}}
\centerline{\includegraphics[clip,scale=0.95]{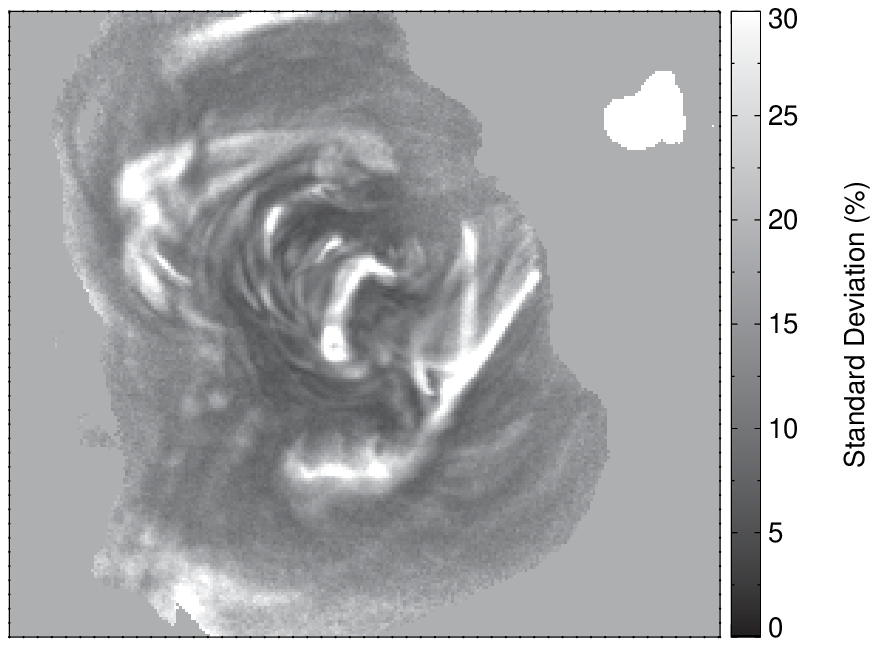}}
\caption{Plots of $\bar{I}$ and $\sigma_I/\bar{I}$ as a function of position for the XRT
  Open/Ti-Poly data. For the $\sigma_I/\bar{I}$ plot only the points with a median
  intensity above 50\,DN s$^{-1}$ are shown. For low intensity pixels the fluctuation
  level is generally very high ($>30$\%).}
\label{fig:xrt_fluct}
\end{figure}

A more systematic view of the variability is obtained by computing the median intensity
($\bar{I}$) and standard deviation ($\sigma_I$) for each point in the co-aligned XRT data
cube. A spatially resolved plot of $\sigma_I/\bar{I}$ is shown in
Figure~\ref{fig:xrt_fluct}. Note we have considered only those pixels with median
intensities above 50\,DN s$^{-1}$. Low intensities well outside the core of the active
region are highly variable.

This calculation shows that there a few structures in the core of the active region that
are highly variable, such as those illustrated in Figure~\ref{fig:xrt}, but that the
emission in most of the active region is constant to within approximately
15\%. \cite{shimizu1995} performed a comprehensive study of active region transient
brightenings with the Soft X-ray Telescope on \textit{Yohkoh}. He found that the
distribution of event energies followed a power-law distribution and that the total energy
in these events was about a factor of 5 smaller than the energy required to heat the
active corona. Our result is qualitatively consistent with this systematic study.

The XRT light curves suggest that the emission in the core of an active region is largely
steady, with relatively few significant brightenings and a very high median intensity
level ($>1000$\,DN s$^{-1}$). This result is consistent with previous results from SXT
\citep[e.g.,][]{porter1995,kano1995}, which showed that the observed emission decayed on
timescales that were long relative to a cooling time.  To put these observations in
perspective, the cooling time for a 50\,Mm loop cooling from equilibrium at 5\,MK is of
the order of 800\,s, while the XRT emission is relatively steady for several hours. Note
that a 50\,Mm loop would connect the middle of the moss regions in this active region.

The core of the active region is clearly made up of the emission from many different loops
and it is unclear if the low fluctuation levels are the result of steady heating on
individual loops or from the superposition of many evolving loops
\citep{cargill1997,cargill2004,patsourakos2006}. Models based on many evolving loops,
however, predict that there should be warm emission ($\sim1$\,MK) that is spatially
correlated with the cooling of hot loops in the core of the active region
\cite[e.g.,][]{warren2007a}. Furthermore, the magnitude of the warm emission should be
related to the observed intensity of the hot emission. Rising levels of hot emission
should be accompanied by a similar increase in the warm emission. These predictions can be
tested easily with the observations of this region with EIS.

\subsection{Warm and Hot Emission}

The interpretation of many active region observations is complicated by the presence of
overlying warm loops. For this region such loops are largely, but not completely, absent,
giving a relatively unobstructed view of the core of the active region at warm
temperatures. This is illustrated by the TRACE 171\,\AA\ movies associated with
Figure~\ref{fig:context}. In these movies we see large scale 1\,MK loops to the north and
south of the active region core but not over the core itself. We also do not see many
loops cooling through the TRACE 171\,\AA\ bandpass that are connected to the moss, similar
to the observations considered by \cite{antiochos2003}.

\begin{figure*}[t!]
\centerline{\includegraphics[clip,scale=0.975]{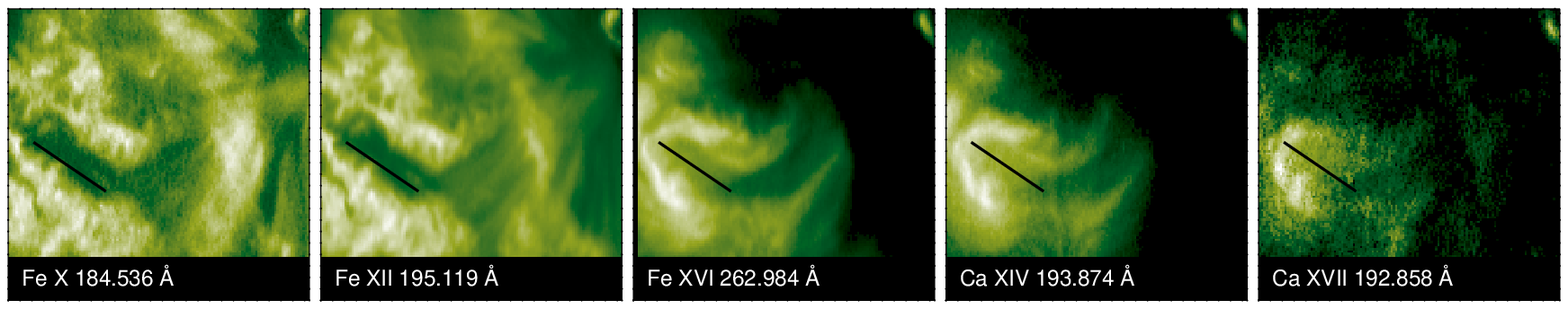}}
\centerline{\includegraphics[clip,scale=0.975]{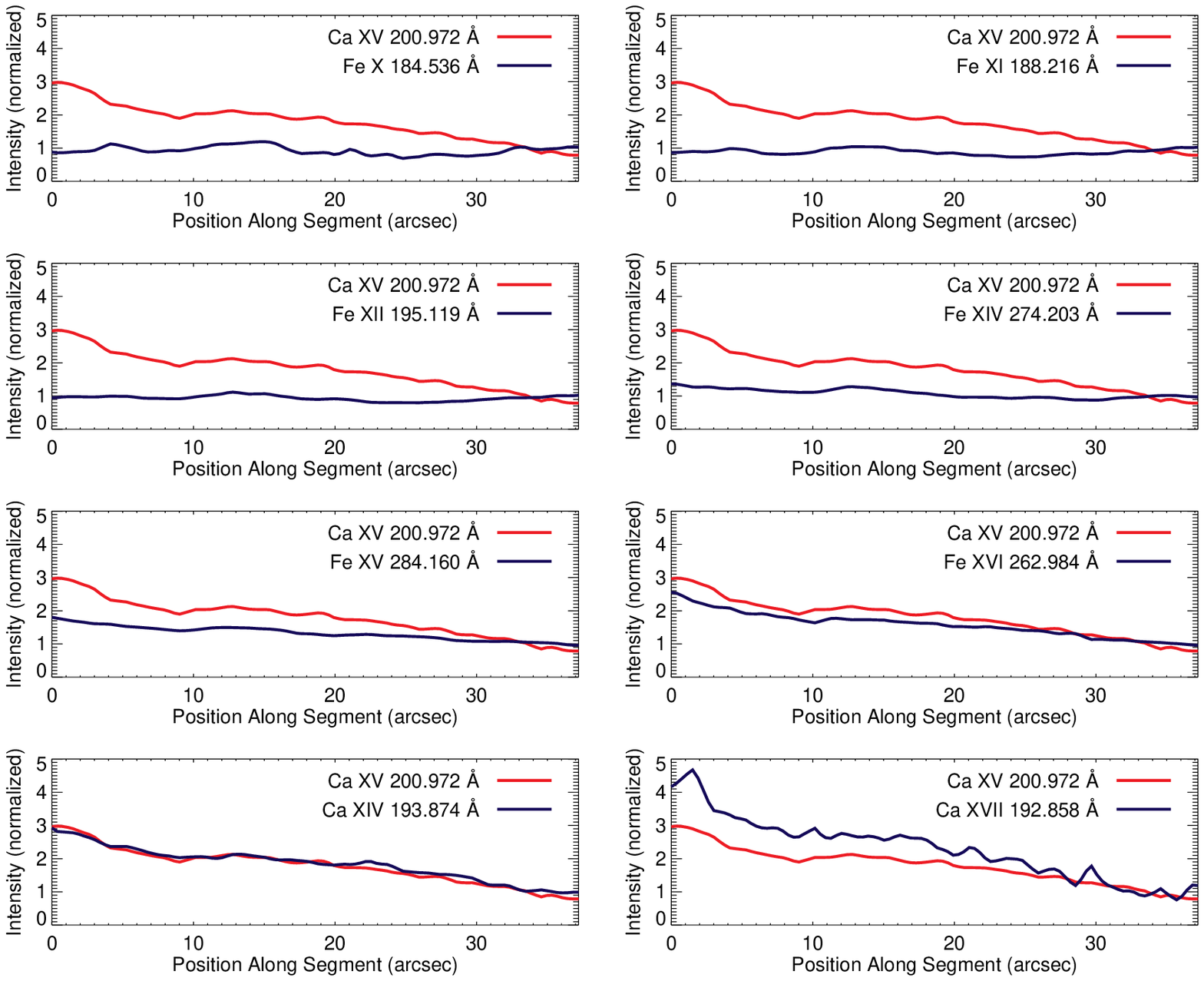}}
\caption{Intensities in the core of an active region. Top panels: EIS rasters in
  \ion{Fe}{10}, \ion{Fe}{12}, \ion{Fe}{16}, \ion{Ca}{14}, and \ion{Ca}{17}.  Bottom
  panels: The intensities along a segment in various emission lines. For each plot the
  intensities are compared with \ion{Ca}{15} 200.972\,\AA. At the lowest temperatures
  (\ion{Fe}{10}--\ion{Fe}{15}) there is little or no relationship between the hot and warm
  emission. At the highest temperatures (\ion{Fe}{16}--\ion{Ca}{17}) the intensities are
  strongly correlated. The intensities are normalized to the values at the end of the
  segment ($\sim40\arcsec$).}
\label{fig:warm_hot}
\end{figure*}

The limited amount of overlying emission allows us to look for any faint warm emission
that is related to the formation of hot coronal loops. Another useful aspect of these
observations is that there is a strong gradient in the intensities of the hot lines
(\ion{Ca}{14}--\textsc{xvii}, \ion{Fe}{17}, and XRT) in the core of the active region in
between the moss. To illustrate the behavior of the emission at different temperatures in
the core of the active region we have selected a line segment parallel to the moss and
extracted the intensities from each of the EIS rasters. These intensities are displayed in
Figure~\ref{fig:warm_hot} where we compare the intensities in each line with those from a
hot line, \ion{Ca}{15} 200.972\,\AA. To facilitate the comparisons we have normalized the
intensities to their values at the end of the line segment where the intensities of the
hot lines are generally the smallest. For the hot lines the intensity along the segment
increases by a factor of 3 to 4 in this region. At the cooler temperatures
(\ion{Fe}{10}--\textsc{xiv}), however, the intensities are unchanged and show no spatial
correlation with the increase of the hot emission. The behavior of the emission from
\ion{Fe}{15} and \ion{Fe}{16} appears to fall in between the hot and warm lines, with
\ion{Fe}{16} following \ion{Ca}{15} fairly closely and \ion{Fe}{15} behaving more like a
warm line.

Since the EIS raster lacks temporal information it is possible to argue that the more
intense emission in the hot lines is evolving and will show up at a later time as enhanced
emission at warmer temperatures.  We have, however, already used the XRT observations to
demonstrate that hot emission in the core is constant over many hours
(Figure~\ref{fig:xrt}). Inspection of later XRT data for this active region shows that the
strong gradients in the intensity and the dark inter-moss region actually persist for
several days (see \citealt{brooks2009b} Figures 1 and 3). The hot emission shown in the
EIS rasters is clearly not dominated by transient events.

We note that the intensities of the warm lines are non-zero in the core of the active
region. Inspection of the TRACE image shown in Figure~\ref{fig:context} suggests that
while the inter-moss region is the dimmest part of the active region core, it is not the
dimmest part of the active region at these temperatures. As we will show in the next
section, almost all of the warm lines have inter-moss intensities that are less than 20\%
of the moss intensities. The \ion{Fe}{13} 202.044\,\AA\ line, however, is unusually bright
in the inter-moss region. The lower level of the transition responsible for this line is
very sensitive to collisonal de-excitation into other levels and the emissivity for this
line rises sharply with decreasing density. This sensitivity to low density background
plasma may account for the enhanced intensity of the \ion{Fe}{13} 202.044\,\AA\ line in
the inter-moss region.

The origin of the warm emission in the core of the active region is unclear. It is most
likely related to unresolved, high lying loops that form around the active region
\citep[e.g.][]{mason1999}. Scattered light from the bright moss may also influence the
observed intensities in this region. See, for example, \cite{deforest2009} for a
discussion of scattered light in the \textit{TRACE} instrument. Scattered light levels
have not yet been measured in EIS, but because EIS and TRACE have several similar design
elements, such as a front entrance filter supported by a mesh and multi-layer coatings,
observations of dim regions with EIS are also likely to suffer from some level of stray
light.

\section{Modeling the Moss}

The moss represents the footpoints of high temperature loops and offers important boundary
conditions for physical models. Observations of the moss are particularly useful for
constraining physical models since the observed intensity is proportional to the loop
pressure and independent of the loop length \citep{martens2000,vourlidas2001}. That is,
\begin{equation}
I_\lambda \propto P_0\times f
\label{eq:ints}
\end{equation}
where $I_\lambda$ is the observed intensity, $P_0$ is the base pressure, and $f$ is the
filling factor. This allows for the calculation of physical models of the moss without
knowing the loop length. This is important since the magnetic topology of an active region
is difficult to infer, even with the use of vector magnetograms
\citep[e.g.,][]{derosa2009}.

Some previous work comparing EIS moss intensities with steady heating models has been
presented in an earlier paper \citep{warren2008}. This work introduced the use of density
sensitive lines to determine both the base pressure and the filling factor. For lines
formed close to \ion{Fe}{12} there was good agreement between steady heating models and
the observations. At the lowest temperatures, however, there was a dramatic difference,
with the modeled intensities being about 400\% higher than what was observed. A similar
temperature dependence in the contrast between moss and network intensities was noted by
\cite{fletcher1999}. Their work suggested that it was difficult to identify moss regions
in relatively low temperature emission.

Here we investigate the possibility that variations in the loop cross section are
responsible for the relatively low moss intensities at low temperatures. For some time it
has been recognized that magnetic flux tubes must expand rapidly in the region between the
high beta photosphere and the low beta corona. That is, coronal loops must look like
``funnels'' in the transition region. This loop expansion plays an important role in
determining the energy balance on coronal loops
\citep[e.g.,][]{gabriel1976,dowdy1987,rabin1991} and needs to be accounted for in modeling
the moss intensities. The network model of \cite{gabriel1976} shows a significant
constriction at temperatures below about 1\,MK, qualitatively similar to what has been
suggested by the EIS observations of the moss.

\begin{figure}
\centerline{\includegraphics[clip,scale=0.98]{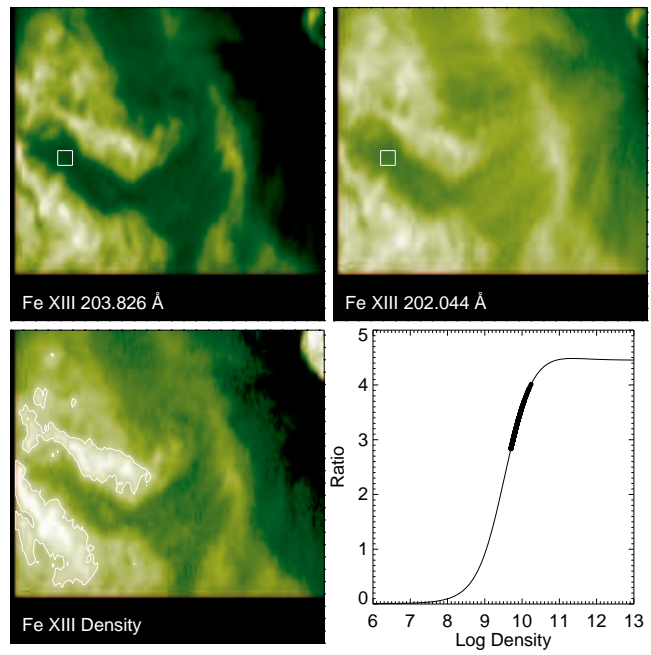}}
\caption{The moss identified using a simple threshold with the density derived from the
  \ion{Fe}{13} 203.826/202.044\,\AA\ ratio. The contour is for $\log n_e = 9.7$. The box
  indicates the region used to estimate the background contribution to the line
  intensities. For each emission line considered here we have computed the median
  intensity in the moss and background regions.}
\label{fig:moss}
\end{figure}

The first step in performing detailed comparisons between theory and observation with
these data is to accumulate the observed moss intensities from this active region. As
shown in Figure~\ref{fig:moss}, we follow \cite{warren2008} and use a density sensitive
line ratio to identify the dense moss regions. For this work we focus on the \ion{Fe}{13}
203.826/202.044\,\AA\ ratio. The threshold has been set at $\log n_e = 9.7$ so that
density contours encompass the bright region in the core of the active region. For each
line we have extracted the intensities in this region and computed the distribution of
intensities. The resulting distribution is approximately Gaussian and for each line we
have determined the median intensity as well as the standard deviation in the distribution
of intensities. A summary of all of the observed moss intensities is given in
Table~\ref{table:ints}.

To account for any contamination from overlying loops or scattered light, we have
subtracted an approximate background derived from the low intensity pixels in the
inter-moss region. For consistency, the same background subtraction procedure is applied
to all of the emission lines. The background intensities are given in
Table~\ref{table:ints}.

It is unclear what uncertainties should be associated with these median intensities. The
moss intensities are relatively high and the statistical errors in the fits to the
Gaussian line profiles are generally small. Furthermore, for simplicity we will work with
the median moss intensities and the uncertainties in these values are also small. Finally,
the errors in our analysis are undoubtedly dominated by the systematic errors in the
atomic data and by the assumptions embedded in the hydro modeling, such the assumption of
ionization equilibrium. Absent any compelling alternative we assume that the uncertainty
in each measured intensity is 15\%, which is much higher than the statistical uncertainty.

Note that in a previous paper we used the \ion{Fe}{12} 186.880/195.119 line pair to
identify the moss and determine the pressure and filling factor \citep{warren2008}. Recent
work, however, has shown that there are discrepancies between the various \ion{Fe}{12} and
\ion{Fe}{13} density sensitive line ratios \citep{young2009b}, leading one to wonder which
densities are the most accurate. As we will discuss in more detail in the next section,
the \ion{Si}{10} 258.375/261.058\,\AA\ ratio is generally consistent with the results from
\ion{Fe}{13} and we use the \ion{Fe}{13} ratio here.

If we assume a constant loop cross section and steady, uniform heating, the only remaining
parameters needed to solve the hydrodynamic loop equations are the volumetric heating rate
and the loop length. As indicated by Equation~\ref{eq:ints}, any family of solutions with
the same base pressure will yield the same observed moss intensity. To illustrate this we
calculate a grid of solutions that covers a range of loop lengths and heating rates. To
solve the hydrostatic loop equations we use a numerical code written by Aad van
Ballegooijen (e.g., \citealt{schrijver2005}). We consider total loop lengths in the range
$L = 10$--100\,Mm and heating rates that yield maximum temperatures in the range $\log
T_{max}=2.5$--7.5\,MK. In the numerical model the lower boundary condition is set at
0.02\,MK and the loops are assumed to be oriented perpendicular to the solar surface.

\begin{figure}
\centerline{\includegraphics[clip,scale=1.0]{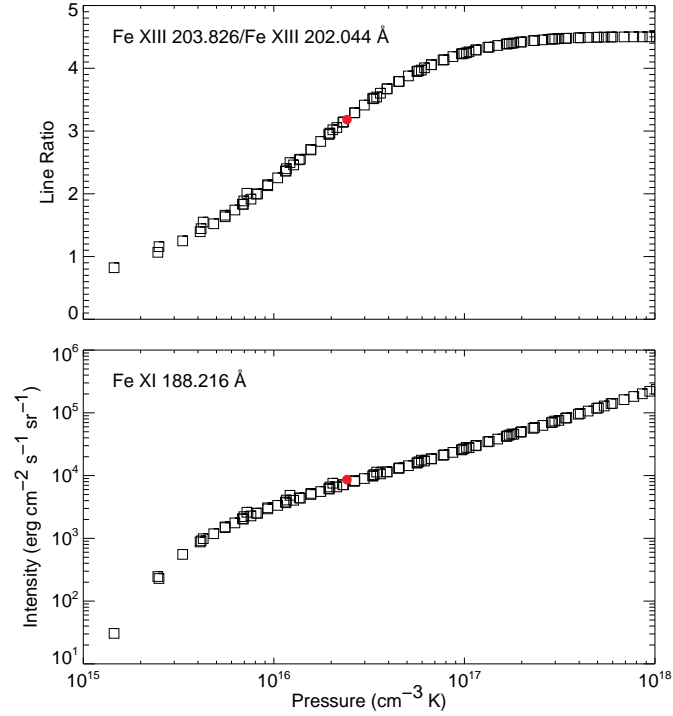}}
\caption{Top panel: The \ion{Fe}{13} 203.826/202.044\,\AA\ ratio as a function of the
  pressure at 1.5\,MK calculated from a steady heating model. Solutions over a range of
  loop lengths and volumetric heating rates have been used. The red dot represents the
  pressure corresponding to the median intensities given in
  Table~\protect{\ref{table:ints}}. Bottom Panel: The intensity in the \ion{Fe}{11}
  188.216\,\AA\ line as a function of pressure.}
\label{fig:moss_pressure}
\end{figure}

For each numerical solution we calculate the total intensity in the line using the
usual expression
\begin{equation}
I_\lambda = \frac{1}{4\pi}\int \epsilon_\lambda(T_e,n_e)n_e^2\,ds,
\end{equation}
where $\epsilon_\lambda(T_e,n_e)$ is the emissivity computed from the CHIANTI atomic
physics database version 5.2.1 \citep{dere1997,landi2006}. Figure~\ref{fig:moss_pressure}
shows the calculated 203.826/202.044\,\AA\ ratio as a function of the pressure at 1.5\,MK,
the temperature of formation for \ion{Fe}{13}. The ratio is clearly a function of the
pressure alone and allows us to easily convert the observed line ratio into a value for
the base pressure. An example calculation for the \ion{Fe}{11} 188.216\,\AA\ intensity is
also shown in Figure~\ref{fig:moss_pressure}. With the pressure determined we can now read
off the calculated intensities for each line. For \ion{Fe}{11} 188.216\,\AA, for example,
the calculated intensity is 8625\,erg cm$^{-2}$ s$^{-1}$ sr$^{-1}$, which is much larger
than what is observed and indicates that the emission is not resolved at the spatial
resolution of EIS.

\begin{figure}
\centerline{\includegraphics[clip,scale=0.5]{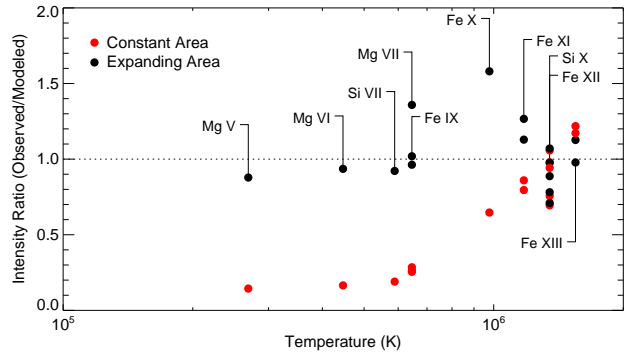}}
\caption{The ratio of observed to calculated intensity for the expanding area (or funnel)
  model and constant area models (const2) plotted as a function of temperature. This plot
  illustrates the failure of the constant cross-section models at low temperatures.}
\label{fig:table_ints}
\end{figure}

Calculated intensities for all of the emission lines are given in
Table~\ref{table:ints}. We have estimated a filling factor of 20\% using the median ratio
of the observed to modeled intensities for the \ion{Fe}{10}, \textsc{xi}, \textsc{xii},
and \textsc{xiii} lines. This corresponds to the \ion{Fe}{11} 188.216\,\AA\ line. For the
emission lines from these ions there is generally good agreement between the observed and
modeled intensities. At lower temperatures, however, the observed intensities are much
lower than what is calculated from the simulation. Ratios of the observed to modeled
intensities as a function of temperature of formation are displayed in
Figure~\ref{fig:table_ints}. This plot shows that the constant area model breaks down at
temperatures below about 1\,MK. This discrepancy is consistent with the comparisons
presented previously \citep{warren2008}.

An alternative approach to modeling the observed emission is to assume a loop length and
determine the heating rate and filling factor that best fits the observed
intensities. This method is based on the assumption that the observed moss intensities are
independent of the loop length, an assumption supported by theory \citep{martens2000} and
by the numerical solutions presented here. This method has the advantage of being
extensible. Our goal is to consider loops with complex geometries and the
intensity-pressure approach outlined in the previous paragraphs becomes very cumbersome
when applied to models with several free parameters. We have implemented such an approach
using the Levenberg-Marquardt least-squares minimization routine \verb+MPFIT+ for the
Interactive Data Language (IDL). The resulting modeled intensities are presented in
Table~\ref{table:ints} and are very similar to those derived from the intensity-pressure
relationships. We have assumed a loop length of 50\,Mm and obtain best-fit parameters of
$\epsilon_0 = 5.98\times10^{-3}$\,erg~cm$^{-3}$~s$^{-1}$ and $f=0.22$. Only the emission
lines formed at \ion{Fe}{10} and above are considered in the minimization. In this
approach all of the lines of interest are considered simultaneously and the pressure
indicated by the \ion{Fe}{13} lines is not necessarily reproduced by the model. To
emphasize these lines in the final solution we reduce the uncertainties assumed for
these lines to 5\%.

To explore the role of geometry in determining the observed intensities we parametrize
the cross-sectional area as a function of height using the following expression
\begin{equation}
A(s) = \frac{1}{\Gamma}\left\{ 1 + \frac{\Gamma - 1}{2}
                \left[\tanh\left(\frac{s-s_0}{\sigma_s}\right)+1\right]\right\},
\end{equation}
for $s<L/2$. The area is extended to the other footpoint by symmetry. The area expansion
is defined so that the minimum is $A(s=0) \approx 1/\Gamma$ and the maximum is $A(s=L/2)
\approx 1$. The gradient in area expansion peaks at $s_0$ and the magnitude of the
gradient is inversely proportional to $\sigma_s$. Smaller values of $\sigma_s$ indicate a
sharper transition. This form has been assumed because it allows for the loop expansion to
be localized. Previous work has considered functions of the form $A\sim T^\alpha$, where
the expansion continues throughout the corona \citep[e.g.,][]{rabin1991,chae1998}. There
is considerable empirical evidence that coronal loops have constant cross sections
\cite[e.g.,][]{klimchuk2000} and our area function is consistent with this. Note that in
our prescription the loop geometry is fixed and does not respond to heating within the
loop.

The expectation is that the constriction will reduce the calculated intensities at the low
temperatures. The discrepancy between the observed intensities appears to occur around the
temperature of formation of \ion{Fe}{10}, which is approximately 1\,MK. In the constant
cross-section case this temperatures occurs at about 0.5\,Mm up the loop length.

\begin{figure}[t!]
\centerline{\includegraphics[clip,scale=0.5]{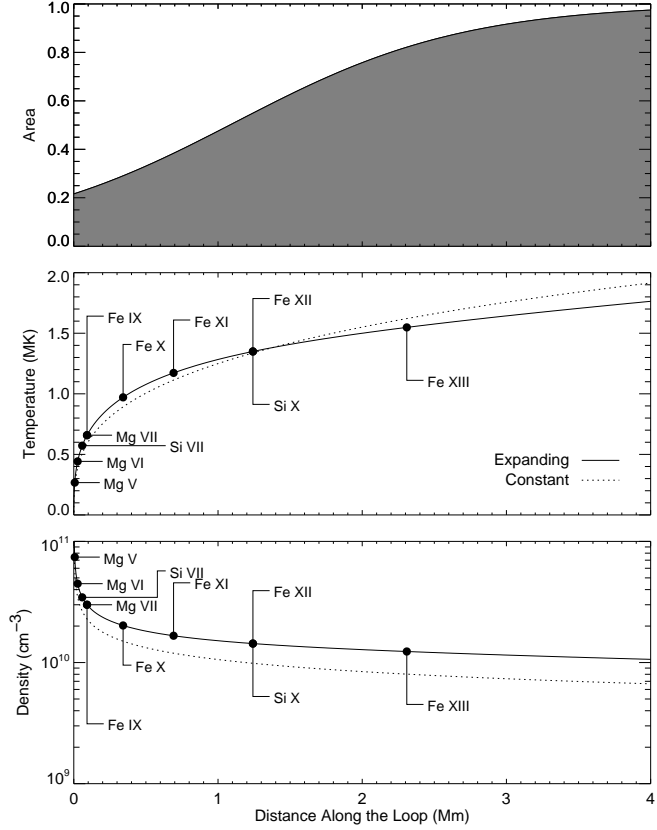}}
\caption{The transition region loop geometry derived from EIS observations of the
  Moss. Top panel: The normalized area as a function of distance along the loop. The
  best-fit parameters are $\epsilon_0 = 8.03\times10^{-3}$, $f=0.11$, $\Gamma=45.6$, $s_0
  = 1.1$\,Mm, and $\sigma_s = 1.6$\,Mm. Bottom panels: The temperature and density along
  the loop. The peak temperature of formation for each ion is indicated on the plots. For
  comparison, the corresponding best-fit solution for the constant cross section case is
  also plotted. The total loop length is 50\,Mm and only part of the solution is shown.}
\label{fig:expand}
\end{figure}

To solve for the best-fit parameters we assume a loop length of 50\,Mm and use the
Levenberg-Marquardt approach. The free parameters are the volumetric heating rate
($\epsilon_0$), the filling factor ($f$), and the loop expansion parameters $\Gamma$,
$s_0$, and $\sigma_s$. An example solution is shown in Figure~\ref{fig:expand} and the
corresponding calculated intensities are given in Table~\ref{table:ints}. The solutions to
the constant cross section and funnel models are generally similar. The constant cross
section model has somewhat higher temperatures and lower densities. The most significant
impact on the calculated intensities comes from the area term in the emission measure
\begin{equation}
EM = A(s)n_e^2\,ds.
\end{equation}
The funnel model clearly reproduces the intensities at the lower temperatures much better
than the constant cross-section models do. In this fit all of the emission lines are
used. To emphasize the \ion{Fe}{13} lines over the other lines the error assumed for the
\ion{Fe}{13} intensities is again reduced to 5\%.

\begin{figure}[t!]
\centerline{\includegraphics[clip,scale=0.5]{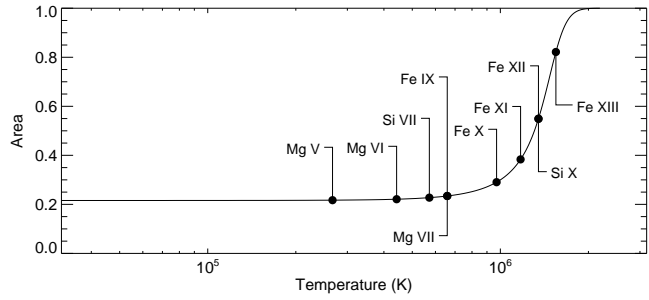}}
\caption{The area as a function of temperature for the expanding loop model shown in
  Figure~\protect{\ref{fig:expand}}. The peak temperature of formation
  for each ion is indicated.}
\label{fig:expand2}
\end{figure}

Previous studies on the transition region geometry \citep[e.g.,][]{rabin1991,chae1998}
have considered the area as a function of temperature and for comparisons with these works
we show $A(T(s))$ derived from the model parameters and the solution to the loop equations
in Figure~\ref{fig:expand2}. Our result is qualitatively similar to the most promising
funnel model derived by \citet[ see model ``B4HL'' in their Figure 12]{rabin1991}. In this
model the area is relatively constant until $\log T \approx 5.6$ and then expands rapidly,
similar to what we show in Figure~\ref{fig:expand}. Our result, however, is not consistent
with the model derived by \cite{chae1998} from SUMER Doppler shift measurements (see their
Figure 19). Their model shows significant expansion at temperatures above $\log T \approx
5.0$. In both cases, as the authors were well aware, comparisons were made with spatially
averaged quiet Sun observations and the applicability of these results to our modeling of
the active region moss is unclear.  \cite{patsourakos1999} used spatially resolved
measurements of the quiet network to examine loop expansion from the transition region to
the corona. The amount of loop expansion that they measured is generally similar to what
we have determined here, although it appears to begin at a somewhat lower temperature than
in our model.

One factor that we have not considered is the impact of chromospheric absorption on the
moss intensities. \cite{depontieu1999} have shown that the evolving, reticulated pattern
of the moss is often related to obscuration by chromospheric jets on neighboring field
lines.  Absorption leads to reduced observed intensities and impacts the filling factors
and heating rates determined from the observations. This chromospheric absorption could
also be stronger at the lowest heights and could explain the lower intensities of the
lower temperature lines. This could also potentially complicate the inference of the
transition region geometry from the observed intensities. Recent analysis of coordinated
EIS and SUMER observations suggests that emission below the Lyman continuum is reduced by
about a factor of 2 \citep{depontieu2009b}. It should be noted that this result is
predicated on comparisons between 2 \ion{Fe}{12} line ratios (186.880/195.119\,\AA\ and
1241.990/195.119\,\AA). As we have discussed, the atomic data for \ion{Fe}{12} is
problematic and there is some uncertainty in this correction. The results from
\ion{Fe}{12} also do not yield insights into the possibility of additional absorption at
lower heights. Clearly more analysis is needed on this important issue.

\section{Electron Densities}

In this paper we have used the \ion{Fe}{13} 203.826/202.044\,\AA\ line ratio to determine
the electron density while previously the \ion{Fe}{12} 186.880/195.119\,\AA\ ratio was
used. The \ion{Fe}{12} ratio is sensitive over a very wide range of densities ($\log n_e =
$ 7 -- 12), making it a very useful plasma diagnostic. It has been found, however, that
the \ion{Fe}{13} and \ion{Fe}{12} densities generally don't agree \citep{young2009b},
leading to the potential for systematic errors in the modeling.

In many previous studies the atomic data for Si has proven to be very robust
\citep[e.g,][]{feldman1999a,doschek1997b} and so we have compared the densities derived
from the \ion{Si}{10} 258.375/261.058\,\AA\ ratio with those from the other line
pairs. The \ion{Si}{10} ratio is sensitive over a much smaller range of densities than the
\ion{Fe}{12} and \ion{Fe}{13} pairs so we have looked for observations of all three line
pairs in small active regions, bright points, and the quiet Sun, i.e., regions where the
densities are generally lower than in the moss region that we have studied here.

\begin{figure*}[t!]
\centerline{\includegraphics[clip,scale=1.0]{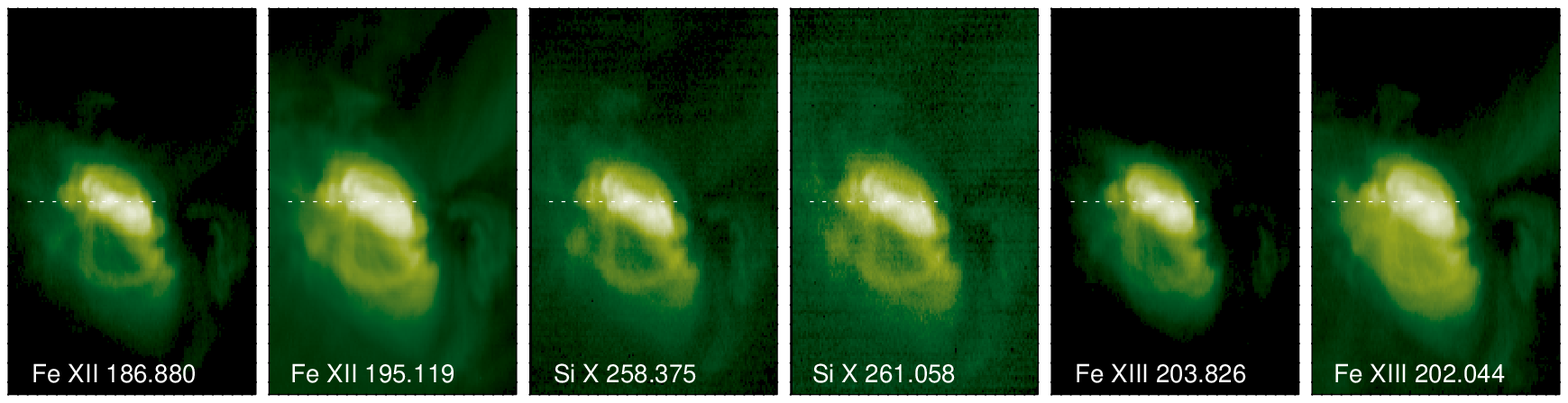}}
\centerline{\includegraphics[clip,scale=1.0]{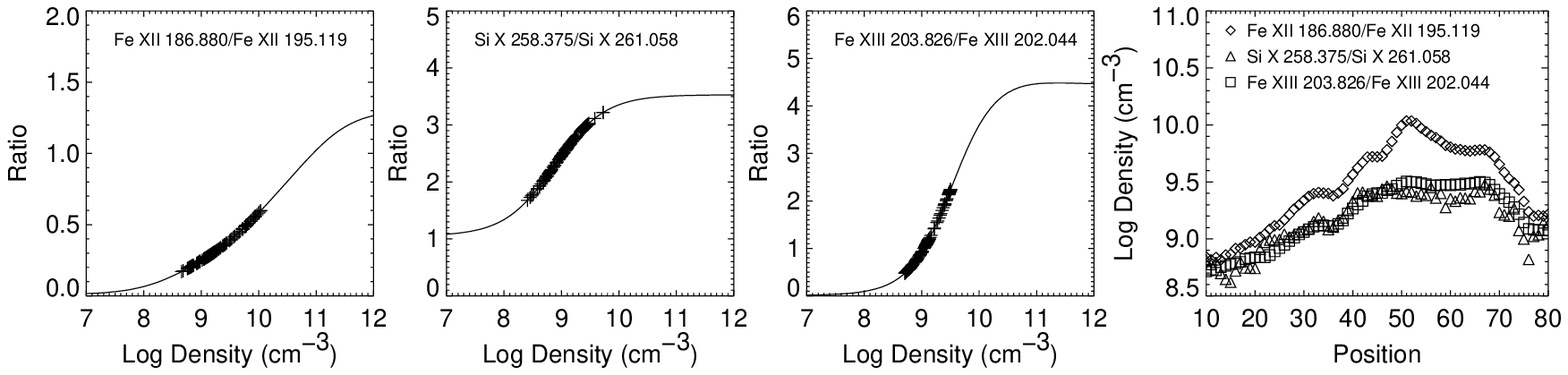}}
\caption{EIS observations of a small active region beginning on 2 December 2007 at
  11:10:44 UT.  Top panels: Rasters in 3 pairs of density sensitive lines. Intensities
  have been extracted from the region indicated by the white line. Bottom panels:
  Densities computed from the 3 pairs of lines. The final panel shows the densities as a
  function of position. The \ion{Fe}{13} and \ion{Si}{10} densities are in generally good
  agreement while the \ion{Fe}{12} densities are systematically higher. }
\label{fig:dens}
\end{figure*}

The result of one such comparison is shown in Figure~\ref{fig:dens}. This plot shows the
densities computed in a small active region from all three line pairs. These calculations
clearly show that at high densities the \ion{Fe}{12} results are systematically higher
than those from \ion{Fe}{13} and \ion{Si}{10} and that the \ion{Fe}{13} and \ion{Si}{10}
densities are in good agreement. We have found similar results from the analysis of 4
other quiet Sun and active region observations. 

The problems with \ion{Fe}{12} become more pronounced at the highest densities, which is
particularly problematic for the analysis of the moss. In these observations the peak
density is about a factor of 3 higher in \ion{Fe}{12}. Since the observed intensity scales
with the square of the density this discrepancy is amplified. We have derived filling
factors of 10--20\% for this moss region. Emphasizing the \ion{Fe}{12} ratio would lead to
filling factors almost an order of magnitude smaller.

\section{Summary}

We have presented a comprehensive analysis of observations in the core of an active region
using data from the EIS and XRT instruments on \textit{Hinode} and \textit{TRACE}. The
apparent steadiness of the XRT emission, the lack of spatial correlation between the hot
and warm emission, and the consistency of the funnel models with the observed emission all
point to frequent heating events that keep the hot loops close to equilibrium.
Furthermore, these results are consistent with high cadence EIS measurements of moss
intensities, Doppler shifts, and nonthermal widths that show little evidence of dynamical
events over many hours \citep{brooks2009b}. In combination, these results provide strong
evidence that the heating in the core of an active region is effectively steady, that is,
the time between heating events is short relative to the relevant radiative and conductive
cooling times.


\acknowledgments Hinode is a Japanese mission developed and launched by ISAS/JAXA, with
NAOJ as domestic partner and NASA and STFC (UK) as international partners. It is operated
by these agencies in co-operation with ESA and NSC (Norway). \textit{TRACE} is supported
by a NASA contract to LMATC. The authors would like to thank Jim Klimchuk for helpful
discussions on the coronal heating problem, George Doschek for an explanation of the
\ion{Fe}{13} 202.044\,\AA\ emissivity, and the referee for a number of very helpful
comments on the original manuscript.



\clearpage

\clearpage

\clearpage

\clearpage

\clearpage

\clearpage

\clearpage

\clearpage

\clearpage

\clearpage

\clearpage

\begin{deluxetable}{lrrrrrrrrrr}
\tabletypesize{\footnotesize}
\tablewidth{170mm}
\tablecaption{Observed and Modeled Moss Intensities\tablenotemark{a}}
\tablehead{
\multicolumn{1}{c}{Ion} &
\multicolumn{1}{c}{$I_{median}$} &
\multicolumn{1}{c}{$\sigma_I$} &
\multicolumn{1}{c}{$I_{back}$} &
\multicolumn{1}{c}{$I_{observed}$} &
\multicolumn{1}{c}{$I_{const1}$} &
\multicolumn{1}{c}{$R$} &
\multicolumn{1}{c}{$I_{const2}$} &
\multicolumn{1}{c}{$R$} &
\multicolumn{1}{c}{$I_{funnel}$} &
\multicolumn{1}{c}{$R$}
}
\startdata
\ion{Mg}{5} 276.579 & 36.4 & 12.4 & 3.1 & 33.3 & 178.3 & 0.19 & 231.3 & 0.14 & 38.0 & 0.88 \\
\ion{Mg}{6} 270.394 & 97.8 & 25.9 & 8.5 & 89.3 & 420.1 & 0.21 & 541.3 & 0.16 & 95.4 & 0.94 \\
\ion{Mg}{7} 280.737 & 96.7 & 27.5 & 5.1 & 91.6 & 247.1 & 0.37 & 321.4 & 0.28 & 67.4 & 1.36 \\
\ion{Si}{7} 275.352 & 136.6 & 42.0 & 9.6 & 126.9 & 529.6 & 0.24 & 668.2 & 0.19 & 137.7 & 0.92 \\
\ion{Fe}{9} 188.497 & 248.0 & 64.6 & 13.8 & 234.2 & 710.1 & 0.33 & 892.6 & 0.26 & 229.9 & 1.02 \\
\ion{Fe}{9} 197.862 & 106.1 & 26.1 & 8.0 & 98.0 & 310.3 & 0.32 & 384.8 & 0.25 & 101.8 & 0.96 \\
\ion{Fe}{10} 184.536 & 936.9 & 225.3 & 62.3 & 874.6 & 1107.0 & 0.79 & 1352.1 & 0.65 & 553.3 & 1.58 \\
\ion{Fe}{11} 182.167 & 1114.6 & 266.3 & 59.0 & 1055.6 & 1114.1 & 0.95 & 1326.6 & 0.80 & 935.0 & 1.13 \\
\ion{Fe}{11} 188.216 & 1827.7 & 398.5 & 121.4 & 1706.4 & 1706.4 & 1.00 & 1984.4 & 0.86 & 1347.6 & 1.27 \\
\ion{Si}{10} 258.375 & 1390.2 & 301.4 & 65.4 & 1324.8 & 1171.6 & 1.13 & 1354.4 & 0.98 & 1355.8 & 0.98 \\
\ion{Si}{10} 261.058 & 455.0 & 97.3 & 24.7 & 430.3 & 355.0 & 1.21 & 406.9 & 1.06 & 402.0 & 1.07 \\
\ion{Fe}{12} 186.880 & 3026.4 & 676.5 & 130.8 & 2895.6 & 2608.4 & 1.11 & 3072.6 & 0.94 & 3263.0 & 0.89 \\
\ion{Fe}{12} 192.394 & 1309.5 & 266.7 & 81.9 & 1227.5 & 1443.1 & 0.85 & 1622.4 & 0.76 & 1571.6 & 0.78 \\
\ion{Fe}{12} 195.119 & 4102.9 & 827.4 & 271.6 & 3831.3 & 4883.6 & 0.78 & 5516.1 & 0.69 & 5395.9 & 0.71 \\
\ion{Fe}{13} 202.044 & 2196.5 & 349.7 & 214.8 & 1981.7 & 1457.3 & 1.36 & 1626.5 & 1.22 & 1759.2 & 1.13 \\
\ion{Fe}{13} 203.826 & 6609.6 & 1284.1 & 304.9 & 6304.7 & 4633.7 & 1.36 & 5379.5 & 1.17 & 6446.1 & 0.98 
\enddata
\tablenotetext{a}{In this table wavelengths are in \AA\ and the intensities are in
erg~cm$^{-2}$~s$^{-1}$~sr$^{-1}$. Note that $\sigma_I$ is the Gaussian width for the
distribution of intensities in the moss and not the statistical uncertainty. $I_{back}$
is an estimate of the background emission in the moss. $I_{const1}$ is the computed
intensity for the constant cross-section model computed from the intensity-pressure
relationships, $I_{const2}$ is the constant cross-section model computed from the
least-squares fitting, and $I_{funnel}$ is the best-fit model for the funnel
geometry. $R$ is the ratio of observed to calculated intensities.}
\label{table:ints}
\end{deluxetable}
 
\clearpage
 

\begin{thebibliography}{}

\bibitem[\protect\citeauthoryear{{Antiochos} et~al.}{{Antiochos}
  et~al.}{2003}]{antiochos2003}
{Antiochos}, S.~K., {Karpen}, J.~T., {DeLuca}, E.~E., {Golub}, L.,  \&
  {Hamilton}, P. 2003, \apj, 590, 547

\bibitem[\protect\citeauthoryear{{Aschwanden} et~al.}{{Aschwanden}
  et~al.}{2008}]{aschwanden2008b}
{Aschwanden}, M.~J., {Nitta}, N.~V., {Wuelser}, J.-P.,  \& {Lemen}, J.~R. 2008,
  \apj, 680, 1477

\bibitem[\protect\citeauthoryear{{Aschwanden}, {Schrijver}, \&
  {Alexander}}{{Aschwanden} et~al.}{2001}]{aschwanden2001b}
{Aschwanden}, M.~J., {Schrijver}, C.~J.,  \& {Alexander}, D. 2001, \apj, 550,
  1036

\bibitem[\protect\citeauthoryear{{Brooks} \& {Warren}}{{Brooks} \&
  {Warren}}{2009}]{brooks2009b}
{Brooks}, D.~H.,  \& {Warren}, H.~P. 2009, \apjl, 703, L10

\bibitem[\protect\citeauthoryear{{Cargill} \& {Klimchuk}}{{Cargill} \&
  {Klimchuk}}{1997}]{cargill1997}
{Cargill}, P.~J.,  \& {Klimchuk}, J.~A. 1997, \apj, 478, 799

\bibitem[\protect\citeauthoryear{{Cargill} \& {Klimchuk}}{{Cargill} \&
  {Klimchuk}}{2004}]{cargill2004}
{Cargill}, P.~J.,  \& {Klimchuk}, J.~A. 2004, \apj, 605, 911

\bibitem[\protect\citeauthoryear{{Chae}, {Yun}, \& {Poland}}{{Chae}
  et~al.}{1998}]{chae1998}
{Chae}, J., {Yun}, H.~S.,  \& {Poland}, A.~I. 1998, \apjs, 114, 151

\bibitem[\protect\citeauthoryear{{Cirtain} et~al.}{{Cirtain}
  et~al.}{2007}]{cirtain2007}
{Cirtain}, J.~W., {Del Zanna}, G., {DeLuca}, E.~E., {Mason}, H.~E., {Martens},
  P.~C.~H.,  \& {Schmelz}, J.~T. 2007, \apj, 655, 598

\bibitem[\protect\citeauthoryear{{Culhane} et~al.}{{Culhane}
  et~al.}{2007}]{culhane2007}
{Culhane}, J.~L., et~al. 2007, \solphys, 60

\bibitem[\protect\citeauthoryear{{De Pontieu} et~al.}{{De Pontieu}
  et~al.}{1999}]{depontieu1999}
{De Pontieu}, B., {Berger}, T.~E., {Schrijver}, C.~J.,  \& {Title}, A.~M. 1999,
  \solphys, 190, 419

\bibitem[\protect\citeauthoryear{{De Pontieu} et~al.}{{De Pontieu}
  et~al.}{2009}]{depontieu2009b}
{De Pontieu}, B., {Hansteen}, V.~H., {McIntosh}, S.~W.,  \& {Patsourakos}, S.
  2009, \apj, 702, 1016

\bibitem[\protect\citeauthoryear{{DeForest}, {Martens}, \&
  {Wills-Davey}}{{DeForest} et~al.}{2009}]{deforest2009}
{DeForest}, C.~E., {Martens}, P.~C.~H.,  \& {Wills-Davey}, M.~J. 2009, \apj,
  690, 1264

\bibitem[\protect\citeauthoryear{Dere et~al.}{Dere et~al.}{1997}]{dere1997}
Dere, K.~P., Landi, E., Mason, H.~E., {Monsignori Fossi}, B.~C.,  \& Young,
  P.~R. 1997, \aaps, 125, 149

\bibitem[\protect\citeauthoryear{{DeRosa} et~al.}{{DeRosa}
  et~al.}{2009}]{derosa2009}
{DeRosa}, M.~L., et~al. 2009, \apj, 696, 1780

\bibitem[\protect\citeauthoryear{Doschek et~al.}{Doschek
  et~al.}{1997}]{doschek1997b}
Doschek, G.~A., Warren, H.~P., Laming, J.~M., Wilhelm, K., Lemaire, P.,
  Sch{\"u}hle, U.,  \& Moran, T.~G. 1997, \apj, 482, L109

\bibitem[\protect\citeauthoryear{{Dowdy}, {Moore}, \& {Emslie}}{{Dowdy}
  et~al.}{1987}]{dowdy1987}
{Dowdy}, J.~F., Jr., {Moore}, R.~L.,  \& {Emslie}, A.~G. 1987, \solphys, 112,
  255

\bibitem[\protect\citeauthoryear{{Feldman} et~al.}{{Feldman}
  et~al.}{1999}]{feldman1999a}
{Feldman}, U., {Doschek}, G.~A., {Sch{\"u}hle}, U.,  \& {Wilhelm}, K. 1999,
  \apj, 518, 500

\bibitem[\protect\citeauthoryear{{Fletcher} \& {de Pontieu}}{{Fletcher} \& {de
  Pontieu}}{1999}]{fletcher1999}
{Fletcher}, L.,  \& {de Pontieu}, B. 1999, \apjl, 520, L135

\bibitem[\protect\citeauthoryear{{Gabriel}}{{Gabriel}}{1976}]{gabriel1976}
{Gabriel}, A.~H. 1976, Royal Society of London Philosophical Transactions
  Series A, 281, 339

\bibitem[\protect\citeauthoryear{{Golub} et~al.}{{Golub}
  et~al.}{2007}]{golub2007}
{Golub}, L., et~al. 2007, \solphys, 243, 63

\bibitem[\protect\citeauthoryear{{Handy} et~al.}{{Handy}
  et~al.}{1999}]{handy1999}
{Handy}, B.~N., et~al. 1999, \solphys, 187, 229

\bibitem[\protect\citeauthoryear{{Kano} \& {Tsuneta}}{{Kano} \&
  {Tsuneta}}{1995}]{kano1995}
{Kano}, R.,  \& {Tsuneta}, S. 1995, \apj, 454, 934

\bibitem[\protect\citeauthoryear{{Klimchuk}}{{Klimchuk}}{2000}]{klimchuk2000}
{Klimchuk}, J.~A. 2000, \solphys, 193, 53

\bibitem[\protect\citeauthoryear{{Ko} et~al.}{{Ko} et~al.}{2009}]{ko2009}
{Ko}, Y.-K., {Doschek}, G.~A., {Warren}, H.~P.,  \& {Young}, P.~R. 2009, \apj,
  697, 1956

\bibitem[\protect\citeauthoryear{{Korendyke} et~al.}{{Korendyke}
  et~al.}{2006}]{korendyke2006}
{Korendyke}, C.~M., et~al. 2006, \ao, 45, 8674

\bibitem[\protect\citeauthoryear{{Landi} et~al.}{{Landi}
  et~al.}{2006}]{landi2006}
{Landi}, E., {Del Zanna}, G., {Young}, P.~R., {Dere}, K.~P., {Mason}, H.~E.,
  \& {Landini}, M. 2006, \apjs, 162, 261

\bibitem[\protect\citeauthoryear{{Lundquist}, {Fisher}, \&
  {McTiernan}}{{Lundquist} et~al.}{2008}]{lundquist2008}
{Lundquist}, L.~L., {Fisher}, G.~H.,  \& {McTiernan}, J.~M. 2008, \apjs, 179,
  509

\bibitem[\protect\citeauthoryear{{Martens}, {Kankelborg}, \&
  {Berger}}{{Martens} et~al.}{2000}]{martens2000}
{Martens}, P.~C.~H., {Kankelborg}, C.~C.,  \& {Berger}, T.~E. 2000, \apj, 537,
  471

\bibitem[\protect\citeauthoryear{{Mason} et~al.}{{Mason}
  et~al.}{1999}]{mason1999}
{Mason}, H.~E., {Landi}, E., {Pike}, C.~D.,  \& {Young}, P.~R. 1999, \solphys,
  189, 129

\bibitem[\protect\citeauthoryear{{Patsourakos} \& {Klimchuk}}{{Patsourakos} \&
  {Klimchuk}}{2006}]{patsourakos2006}
{Patsourakos}, S.,  \& {Klimchuk}, J.~A. 2006, \apj, 647, 1452

\bibitem[\protect\citeauthoryear{{Patsourakos} et~al.}{{Patsourakos}
  et~al.}{1999}]{patsourakos1999}
{Patsourakos}, S., {Vial}, J.~C., {Gabriel}, A.~H.,  \& {Bellamine}, N. 1999,
  \apj, 522, 540

\bibitem[\protect\citeauthoryear{{Porter} \& {Klimchuk}}{{Porter} \&
  {Klimchuk}}{1995}]{porter1995}
{Porter}, L.~J.,  \& {Klimchuk}, J.~A. 1995, \apj, 454, 499

\bibitem[\protect\citeauthoryear{{Rabin}}{{Rabin}}{1991}]{rabin1991}
{Rabin}, D. 1991, \apj, 383, 407

\bibitem[\protect\citeauthoryear{{Schrijver} et~al.}{{Schrijver}
  et~al.}{2004}]{schrijver2004}
{Schrijver}, C.~J., {Sandman}, A.~W., {Aschwanden}, M.~J.,  \& {DeRosa}, M.~L.
  2004, \apj, 615, 512

\bibitem[\protect\citeauthoryear{{Schrijver} \& {van Ballegooijen}}{{Schrijver}
  \& {van Ballegooijen}}{2005}]{schrijver2005}
{Schrijver}, C.~J.,  \& {van Ballegooijen}, A.~A. 2005, \apj, 630, 552

\bibitem[\protect\citeauthoryear{{Shimizu}}{{Shimizu}}{1995}]{shimizu1995}
{Shimizu}, T. 1995, \pasj, 47, 251

\bibitem[\protect\citeauthoryear{{Spadaro} et~al.}{{Spadaro}
  et~al.}{2003}]{spadaro2003}
{Spadaro}, D., {Lanza}, A.~F., {Lanzafame}, A.~C., {Karpen}, J.~T.,
  {Antiochos}, S.~K., {Klimchuk}, J.~A.,  \& {MacNeice}, P.~J. 2003, \apj, 582,
  486

\bibitem[\protect\citeauthoryear{{Ugarte-Urra}, {Warren}, \&
  {Brooks}}{{Ugarte-Urra} et~al.}{2009}]{urra2009}
{Ugarte-Urra}, I., {Warren}, H.~P.,  \& {Brooks}, D.~H. 2009, \apj, 695, 642

\bibitem[\protect\citeauthoryear{{Vourlidas} et~al.}{{Vourlidas}
  et~al.}{2001}]{vourlidas2001}
{Vourlidas}, A., {Klimchuk}, J.~A., {Korendyke}, C.~M., {Tarbell}, T.~D.,  \&
  {Handy}, B.~N. 2001, \apj, 563, 374

\bibitem[\protect\citeauthoryear{{Warren}, {Feldman}, \& {Brown}}{{Warren}
  et~al.}{2008}]{warren2008c}
{Warren}, H.~P., {Feldman}, U.,  \& {Brown}, C.~M. 2008, \apj, 685, 1277

\bibitem[\protect\citeauthoryear{{Warren} \& {Winebarger}}{{Warren} \&
  {Winebarger}}{2006}]{warren2006b}
{Warren}, H.~P.,  \& {Winebarger}, A.~R. 2006, \apj, 645, 711

\bibitem[\protect\citeauthoryear{{Warren} \& {Winebarger}}{{Warren} \&
  {Winebarger}}{2007}]{warren2007a}
{Warren}, H.~P.,  \& {Winebarger}, A.~R. 2007, \apj, 666, 1245

\bibitem[\protect\citeauthoryear{{Warren}, {Winebarger}, \&
  {Hamilton}}{{Warren} et~al.}{2002}]{warren2002b}
{Warren}, H.~P., {Winebarger}, A.~R.,  \& {Hamilton}, P.~S. 2002, \apjl, 579,
  L41

\bibitem[\protect\citeauthoryear{{Warren}, {Winebarger}, \& {Mariska}}{{Warren}
  et~al.}{2003}]{warren2003}
{Warren}, H.~P., {Winebarger}, A.~R.,  \& {Mariska}, J.~T. 2003, \apj, 593,
  1174

\bibitem[\protect\citeauthoryear{{Warren} et~al.}{{Warren}
  et~al.}{2008}]{warren2008}
{Warren}, H.~P., {Winebarger}, A.~R., {Mariska}, J.~T., {Doschek}, G.~A.,  \&
  {Hara}, H. 2008, \apj, 677, 1395

\bibitem[\protect\citeauthoryear{{Winebarger}, {Warren}, \&
  {Falconer}}{{Winebarger} et~al.}{2008}]{winebarger2008}
{Winebarger}, A.~R., {Warren}, H.~P.,  \& {Falconer}, D.~A. 2008, \apj, 676,
  672

\bibitem[\protect\citeauthoryear{{Winebarger}, {Warren}, \&
  {Mariska}}{{Winebarger} et~al.}{2003}]{winebarger2003}
{Winebarger}, A.~R., {Warren}, H.~P.,  \& {Mariska}, J.~T. 2003, \apj, 587, 439

\bibitem[\protect\citeauthoryear{{Winebarger}, {Warren}, \&
  {Seaton}}{{Winebarger} et~al.}{2003}]{winebarger2003b}
{Winebarger}, A.~R., {Warren}, H.~P.,  \& {Seaton}, D.~B. 2003, \apj, 593, 1164

\bibitem[\protect\citeauthoryear{{Young} et~al.}{{Young}
  et~al.}{2009}]{young2009b}
{Young}, P.~R., {Watanabe}, T., {Hara}, H.,  \& {Mariska}, J.~T. 2009, \aap,
  495, 587

\end{thebibliography}
\end{document}